\documentclass[traditabstract]{aa} 
\usepackage{graphicx}
\usepackage{natbib}
\usepackage{txfonts}
\begin{document}
\title{About the p-mode frequency shifts in HD~49933}

         \author{D. Salabert
          \inst{\ref{inst:iac},\ref{inst:ul},\ref{inst:un}}\thanks{salabert@oca.eu}
          \and
          C. R\'egulo
          \inst{\ref{inst:iac},\ref{inst:ul}}
          \and
          J. Ballot
          \inst{\ref{inst:irap},\ref{inst:ut}}
          \and
          R.~A. Garc\'ia
          \inst{\ref{inst:aim}}
          \and
          S. Mathur
          \inst{\ref{inst:hao}}
          }

\institute{Instituto de Astrof\'isica de Canarias, E-38200 La Laguna, Tenerife, Spain\label{inst:iac}
         \and
              Departamento de Astrof\'isica,  Universidad de La Laguna, E-38206 La Laguna, Tenerife, Spain\label{inst:ul}
         \and
         	    Universit\'e de Nice Sophia-Antipolis, CNRS, Observatoire de la C\^ote d'Azur, BP 4229, 06304 Nice Cedex 4, France\label{inst:un}
 \and
            Institut de Recherche en Astrophysique et Plan\'etologie, CNRS, 14 av. E. Belin, 31400 Toulouse, France\label{inst:irap}
 \and
            Universit\'e de Toulouse, UPS-OMP, IRAP,  31400 Toulouse, France\label{inst:ut}
         \and
             Laboratoire AIM, CEA/DSM-CNRS, Universit\'e Paris 7 Diderot, IRFU/SAp, Centre de Saclay, 91191 Gif-sur-Yvette, France\label{inst:aim}
         \and
           High Altitude Observatory, NCAR, PO Box 3000, Boulder, CO 80307, USA\label{inst:hao}
}
   
   \date{Received xxxx; accepted xxxx}

 
  \abstract
 {We study the frequency dependence of the frequency shifts of the low-degree p modes measured in the F5V star HD~49933, by analyzing the second run of observations collected by the CoRoT satellite. 
The 137-day light curve is divided into two subseries corresponding to periods of low and high stellar activity. The activity-frequency relationship is obtained independently from the analysis of the mode frequencies extracted by both a local and a global peak-fitting analyses, and from a cross-correlation technique in the frequency range between 1450~$\mu$Hz and 2500~$\mu$Hz. The three methods return consistent results. We show that the frequency shifts measured in HD~49933 present a frequency dependence with a clear increase with frequency, reaching a maximal shift of about 2~$\mu$Hz around 2100~$\mu$Hz. Similar variations are obtained between the $l=0$ and $l=1$ modes. At higher frequencies, the frequency shifts show indications of a downturn followed by an upturn, consistent between the $l=0$ and 1 modes. We show that the frequency variation of the p-mode frequency shifts of the solar-like oscillating star HD~49933 has a comparable shape to the one observed in the Sun, which is understood to arise from changes in the outer layers due to its magnetic activity.}



   \maketitle
  
%

\section{Introduction}
The convective motions in the external layers of  solar-like oscillating stars excite acoustic sound waves, which are trapped in the stellar interiors \citep[e.g.,][]{Goldreich77,Goldreich88}. The precise frequencies of these pressure (p) driven waves depend on the properties of the medium in which they propagate. Thus, helio- and asteroseimology are able to infer the properties of the Sun and stellar interiors by studying and characterizing these p modes \citep[e.g.,][]{Gough96,Metcalfe10b}. In particular, since the very first helioseismic observations, different authors noticed that near the maximum of the solar magnetic activity cycle, the frequency of the modes were shifted towards higher values \citep[e.g.,][]{Woodard85,fossat87,palle89,ang92}. The frequency shifts, which follow an inverse mode-mass scaling \citep{libbrecht90}, are explained to arise from changes in the outer layers of the Sun along the 11-year cycle \citep{Goldreich91,dziem04,dziem05}. The high-frequency p modes, which have higher upper turning points, are then more sensitive to the perturbations induced by the magnetic field in the outer part of the Sun than the low-frequency p modes, that are less sensitive to the outer layers. Thus, the high-frequency p modes are observed to have large variations with activity, while the low-frequency p modes present little or no variation with activity \citep[e.g.,][]{libbrecht90,ronan94,chaplin98,gelly02,salabert04,howe08}. 

As longer, continuous, high-quality helioseismic measurements were collected, the other p-mode oscillation parameters, such as the amplitude, the linewidth, and the asymmetry, were also observed to vary with the solar activity cycle \citep[e.g,][]{Chaplin00,Komm02,salabert03,salabert06,JimenezReyes07}. While the frequency shifts were proven to be closely correlated with the surface activity proxies over the last three 
solar cycles (21, 22, and 23) \citep[e.g.,][]{chaplin07b}, clear differences were recently shown between the global p-mode frequencies and the surface activity of the Sun during the unusually extended minimum of cycle 23 \citep{broomhall09,salabert09,tripathy10}. 
Furthermore, a faster cyclic frequency variations, with a period of about 2 years, was observed to coexist in the Sun \citep{Benevolenskaya95,Fletcher10}, which is believed to have an origin located deeper inside the convective zone.

Magnetic cycles in other stars were already reported by several authors \citep[e.g.,][]{Wilson78,Baliunas85,Baliunas95,Hall07}, and the observed cycles are covering a range between 2.5 and 25~years. Moreover, several authors suggested that the periods of the activity cycles increase proportional to the stellar rotational periods along two distinct paths in main-sequence stars: the active and the inactive stars \citep{Saar99,BV07}. Recently, \citet{Metcalfe10a} uncovered a 1.6-year magnetic activity cycle in the exoplanet host star HD~17051 by measuring the \ion{Ca}{ii}
 H and K emission lines. However, techniques based on spectroscopic measurements are only good proxies of surface magnetic fields \citep{Leighton59}. Spectropolarimetric observations allow direct measurements of stellar magnetic fields, that has revealed, for instance, reversals of magnetic field polarity during stellar cycles for F and G stars \citep[see][]{Donati08, Fares09, Petit09}.

The recent discovery of variations with magnetic activity in the p-mode oscillation frequencies of the F5V star HD~49933 using asteroseismic techniques \citep{garcia10} opens a new era in the study of the physical phenomena involved in the dynamo processes. Indeed, the seismic observables provide invaluable information to pierce inside the stars and determine key parameters for the study of stellar activity, such as the depth of the convection zone, the characteristic evolution time of the granulation, the differential rotation, or the sound-speed as a function of the star's radius. All these new observables will impose new constraints to the theory \citep[e.g.][]{Chaplin07a,Metcalfe07}. Long, high-quality photometric observations being currently collected by the space-based Convection, Rotation, and planetary Transits \citep[CoRoT,][]{baglin06} and {\it Kepler} \citep{Koch10} missions will certainly contribute to these studies \citep[e.g.][]{Barban09,garcia09,Deheuvels10,Mathur10a,Chaplin10}. In this paper, we extend the original analysis from \citet{garcia10} by studying the frequency dependence of the activity-frequency relationship.

\section{Data analysis}
\label{sec:anal}
The light curve of HD~49933 was collected during two observational runs from the CoRoT satellite \citep{app08,benomar09}. The first run was performed in 2007 over a period of 60 days, while in 2008 a second run of observations was obtained over 137 days. In the present paper, we analyzed the data from the second run only as it is the longest run and that it corresponds to an important variation in the magnetic activity of HD~49933 and its p-mode frequencies and amplitudes \citep{garcia10}. These continuous 137 days of observations were divided into two subseries depending on the activity level of HD~49933 as observed by \citet{garcia10}. The first 90 days of the original time series correspond to a period of low activity, and the following 47 days correspond to a period of higher activity. Thus, these two subseries optimize the temporal coverage of the observed magnetic activity as well as the frequency resolution. The frequency dependence of the frequency shifts of HD~49933 were obtained by three independent methods: 1) by means of the mode frequencies extracted by a local (method $\#1$) and global (method $\#2$) peak-fitting procedures, and 2) by using a cross-correlation analysis (method $\#3$). Details of the methods are given below. A total of twelve overtones over the frequency range from 1450~$\mu$Hz up to 2500~$\mu$Hz of the oscillation power spectrum were analyzed.

\subsection{Mode peak-fitting methods}
\label{sec:fitting}
The power spectrum of the two subseries was obtained in order to extract estimates of the p-mode parameters. The mode identification provided by \citet{benomar09} was adopted. The stellar background was modeled using six free parameters as in \citet{Mathur10b} and the individual frequencies from \citet{benomar09} were taken as initial guesses. Both local and global peak-fitting techniques were used. 
The local peak-fitting procedure was performed by fitting sequences of successive series of $l=0$, 1, and 2 modes using a maximum-likelihood estimator (method $\#1$). The individual $l=0$, 1, and 2 modes were modeled using a single Lorentzian profile for each of the angular degrees $l$. Therefore, neither rotational splitting nor inclination angle were included in the fitting model.  The amplitude ratios between the $l=0$, 1, and 2 modes were fixed to 1, 1.5, and 0.5 respectively, and only one linewidth was fitted per radial order. Although the $l=2$ modes are visible in the power spectrum, the accuracy of their frequency measurements is rather small compared to the $l=0$ and 1 modes \citep[see for instance, Table~1 of][]{benomar09}. We decided then to use only the $l=0$ and 1 modes to calculate the frequency shifts from the individual mode frequencies extracted with the method~$\#1$.
Because the very short mode lifetimes in HD~49933, the $l=2$ modes are blended with the neighboring $l=0$ modes, which makes the determination of the $l=0$ and 2 mode frequencies strongly negatively correlated, especially at high frequency where the modes are broader. Due to this strong blending, we decided to determine the centroid of the even $l$ modes by fitting each pair $l=0$ and 2 with a single Lorentzian profile. We then performed a global fitting of the spectrum including the stellar background and about 20 Lorentzian profiles (method $\#2$). The frequencies obtained for the even pairs of modes correspond to a weighted average of the $l=0$ and $l=2$ frequencies, the weights being the relative visibilities of the modes. If we assume: 1) that the relative amplitude between the $l=0$ and $2$ modes does not change either with frequency or with the activity level, and 2) that the frequency shifts depend mainly on the frequency, i.e. that close pairs of $l=0$ and 2 modes have similar activity-induced frequency variations, then the shift measured for the even centroids is equal to the shift of the $l=0$ modes.
Note also that the fitted frequencies for the odd pairs of modes correspond to the $l=1$ frequencies, as the $l=3$ modes have neglecting contributions. 
The mode parameters thus extracted from methods~$\#1$ and $\#2$ were checked to be consistent within the errors with \citet{benomar09}.
For both methods $\#1$ and $\#2$, the frequency shifts $\delta\nu_l$ were obtained by subtracting the individual frequencies measured during the 47-day period of high magnetic activity and the 90-day period of low magnetic activity \citep[similarly to what is performed with the Sun, see e.g.,][]{salabert04}. Additionally, we analyzed the temporal variations of the mode amplitudes and linewidths, but the results are not reliable due to a poor determination of these parameters at the level of precision of the data.

   \begin{figure*}
   \centering
     \includegraphics[width=\textwidth]{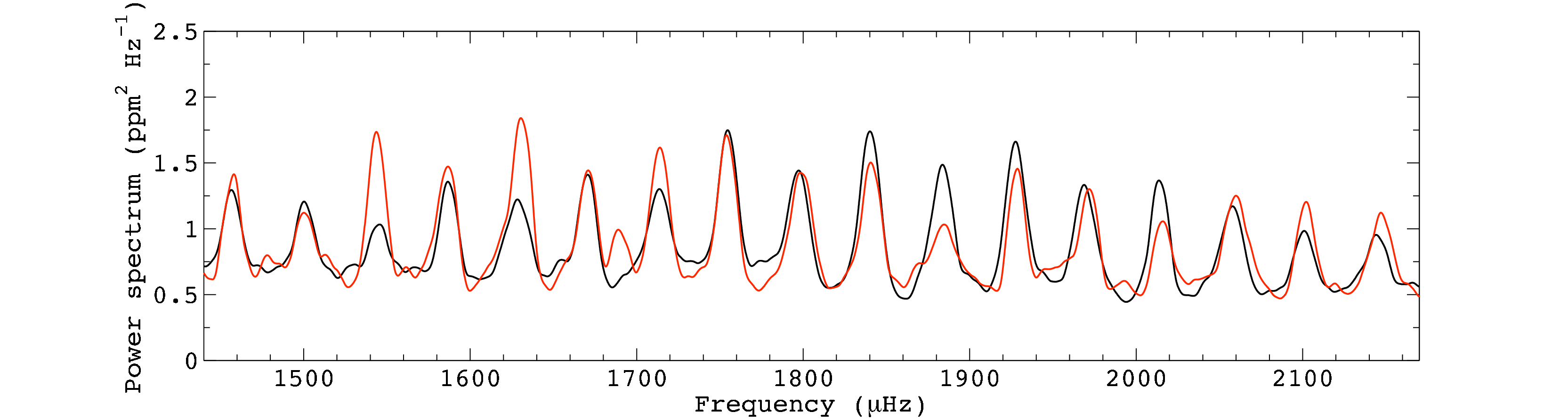}
 \caption{Smoothed power spectra of the 90-day period of low stellar activity (black) and of the 47-day period of high activity (red) of the analyzed observations of HD~49933 represented over the frequency range between 1440 and 2170 $\mu$Hz. }
         \label{fig:fig1}
   \end{figure*}
%

\subsection{Cross-correlation method}
\label{sec:cross}
The cross-correlation method was first used by \citet{palle89} in order to measure the solar frequency shift from one single instrument. In our case, the power spectrum of the 47 days at high activity was correlated with the power spectrum of the 90 days at low activity (method $\#3$). As the two datasets do not have the same number of days, the shortest subseries was zero-padded in order to have the same number of
bins in both power spectra. The background noise was modeled as in Sect.~\ref{sec:fitting}, and subtracted from the power spectra. To obtain the frequency shift, a Gaussian profile was fitted to the cross-correlation function to estimate its maximum, the Gaussian profile being centered at the position given by the skewness of the cross-correlation function to avoid any bias due to asymmetry. Thus, the centroid of the fitted Gaussian returned the amount of shift in frequency, the associated uncertainty being given by the 1$\sigma$ error bar of the fit.
While the local and global peak-fitting methods allow us to study the frequency shifts of the individual modes separately (Sect.~\ref{sec:fitting}), the cross-correlation method returns a global value of the frequency shift
for all the visible modes in the power spectrum. In order to obtain estimates of the frequency shifts for the even and odd pairs of modes, two set of spectra were obtained from each spectra. These new spectra were built by extracting the points containing only the frequency windows of the corresponding odd or even pairs, and creating thus a new array of points. Then, the cross-correlation functions were calculated in the same manner as previously and the associated frequency shifts of the even and odd pairs of modes returned.

\begin{figure*}
\centering
\includegraphics[width=0.7\textwidth]{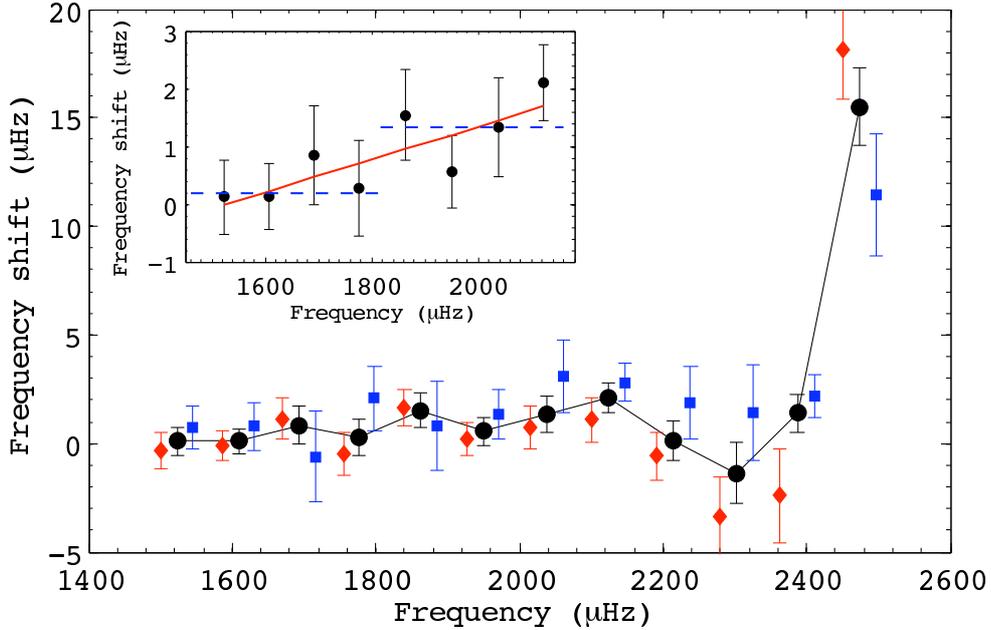}
\caption{Frequency shifts of the $l=0$ (blue squares) and $l=1$ modes (red diamonds) of HD~49933 as a function of frequency obtained from method $\#1$. The corresponding weighted mean values of the frequency shift between consecutive $l=0$ and 1 modes are represented by the black dots (solid line). A zoom-in on the frequency range up to 2100~$\mu$Hz is shown in the insert, which represents the mean frequency shift. The horizontal blue dashed lines correspond to the frequency shifts obtained with the cross-correlation method ($\#3$). The solid red line corresponds to a weighted linear fit.}
\label{fig:fig2}
\end{figure*}

\section{Results}
\label{sec:res}
The power spectra of HD~49933 corresponding to the 90-day period of low stellar activity and the 47-day period of higher activity are shown on Fig.~\ref{fig:fig1} between 1440 and 2170 $\mu$Hz. For illustrative purpose, the spectra were heavily smoothed over 20 $\mu$Hz, which is comparable to twice the mode linewidth at high frequency \citep[see Fig.~4 in][]{benomar09}. Both power spectra overlap in frequency in the low-frequency range below 1780 $\mu$Hz, while a positive shift becomes visible as the mode frequency increases during the period of higher stellar activity. We note the presence of a prominent peak at about 1690 $\mu$Hz in the 47-day power spectrum of high activity. We checked that it does not correspond to a harmonic of the spacecraft orbit, therefore it might have an instrumental origin. Figure~\ref{fig:fig2} shows the frequency shifts of HD~49933 as a function of frequency measured between the periods of high and low activity using the results from the local peak-fitting analysis (method $\#1$). Similar results are obtained with the global peak-fitting analysis (method $\#2$). A frequency dependence of the $l=0$ and 1 p-mode frequency shifts can be observed. 
The mean frequency shifts between consecutive $l=0$ and 1,  using the formal errors as weights, 
are also represented. The frequency shift increases with frequency up to about 2100 $\mu$Hz to reach a maximum of $\sim$ 2 $\mu$Hz, followed by a downturn with a reversal of the sign down to about $-1.5$ $\mu$Hz at $\sim$ 2300~$\mu$Hz for the mean frequency shift. Moreover, this downturn might be followed by an upturn for frequencies above 2300 $\mu$Hz. 
The individual $l=0$ and $l=1$ modes have, within the errors, similar variations with frequency. But the $l=1$ frequency shifts are overall better constrained than the $l=0$ frequency shifts because of the small amplitude of the $l=3$ modes in the power spectrum. Besides, the $l=1$ modes are observed to have a deeper downturn above 2100~$\mu$Hz than the $l=0$ modes.
Nevertheless, these downturn and upturn, although consistent between the $l=0$ and $l=1$ frequency shifts, occur when the SNR of the modes start to get small. Thus we cannot exclude any effects due to the low SNR to explain the behavior of the frequency shifts at high frequency. Longer datasets will be needed to improve the SNR of the measurements in order to confirm the observed shifts at high frequency. 
However, if they are real, these reversals of the sign at high frequencies can be qualitatively explained by changes in chromospheric magnetic field and temperature \citep{johnston95}. The insert in Fig.~\ref{fig:fig2} shows the increasing part of the mean frequency shift 
up to 2100~$\mu$Hz, which would reflect structural changes in and just below the photosphere with stellar activity if we suppose similar mechanisms as in the Sun \citep[see, e.g.,][]{Goldreich91}. The solid red line, which corresponds to a weighted linear fit, clearly shows the increase in frequency shift as a function of frequency. We note the excellent agreement with the results 
from the cross-correlation method (horizontal dashed lines).
We also calculated the variations of the mean mode amplitude as a function of frequency using the same methodology as in \citet{garcia10},  but by separating the power spectrum into two regions. 
But no reliable results were obtained due to the very large uncertainties related to the reduced frequency ranges.

\begin{figure}
\centering
\includegraphics[width=0.47\textwidth]{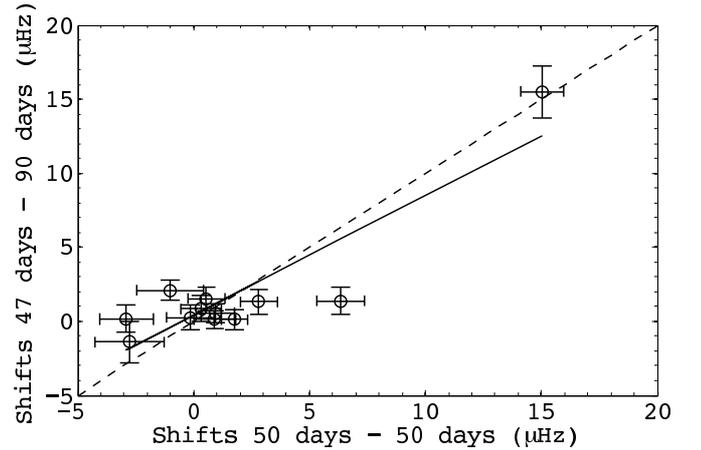}
\caption{Comparison between the frequency shifts obtained from the analysis of the 90-day and 47-day subseries with those from the analysis of the two 50-day subseries in the case of method $\#1$. The solid line corresponds to the linear fit between the data, and the dashed line represents the 1 : 1 correlation.}
\label{fig:fig3}
\end{figure}

Table~\ref{table:fitting} gives the mean frequency shifts obtained from the local analysis of the individual mode frequencies (method $\#1$) averaged between 1470 $\mu$Hz and 1815 $\mu$Hz, and between 1815 $\mu$Hz and 2160 $\mu$Hz, using the formal errors as weights. These frequency ranges were chosen to contain the same number of overtones of $l=0$ and $l=1$ modes, and correspond to the represented frequency shifts in the insert of Fig.~\ref{fig:fig2}. The results obtained over the same frequency ranges from the analyses from the global fitting of the p-mode spectrum (method $\#2$)  and the cross-correlation technique (method $\#3$) are also given in Table~\ref{table:fitting}. The three independent methods agree within the 1$\sigma$ errorbars. The $l=0$ modes -- and the even pair of modes ($l=0,2$) as well -- are observed to have, in average, larger frequency shifts than the $l=1$ modes. The total amount of change in mode frequency between periods of low and high stellar activity of HD~49933 is consistent with the temporal variation of the mode frequencies with magnetic activity measured by \citet{garcia10}. In order to check the consistency of the results, we also analyzed two subseries of 50 days each, centered on the period of low activity and on the rising phase of the magnetic activity. Similar results as Fig.~\ref{fig:fig2} and Table~\ref{table:fitting} were obtained from the mode peak-fitting and the cross-correlation methods. For example, Fig.~\ref{fig:fig3} compares the mean frequency shifts obtained from the analysis of the 90-day and 47-day subseries with those from the analysis of the two 50-day subseries in the case of the local mode peak-fitting analysis (method $\#1$). The associated Pearson's linear correlation coefficient is about 0.89.

The frequency dependence of the frequency shifts measured in HD~49933 is comparable in shape to the one observed in the Sun, using both global \citep[e.g.,][]{libbrecht90,chaplin98,gelly02,salabert04} and local \citep[e.g.,][]{howe08} helioseismic measurements. That indicates then the presence of similar physical phenomena driving the frequency shifts of the modes of oscillation as the ones taking place in the Sun, which are understood to arise from changes in the outer layers of the star along its activity cycle. However, the frequency shift measured in HD~49933 is at least five times larger than in the Sun, which is of about 0.5 $\mu$Hz at 3700 $\mu$Hz between the maximum and the minimum of the 11-year solar cycle \citep[e.g.][]{salabert04}.

\begin{table}
\caption{Mean frequency shifts $\langle \delta\nu_l \rangle$ in $\mu$Hz over two frequency ranges obtained from the local and global mode peak-fitting, and the cross-correlation analyses for HD~49933.}            
\label{table:fitting}      
\centering                        
\begin{tabular}{cccc}        
\hline\hline                
Method &  $\langle \delta\nu_l \rangle$   &  \multicolumn{2}{c}{Frequency range}\\
 &           &  \multicolumn{2}{c}{($\mu$Hz)}\\
 & 	      & [1470--1815] & [1815--2160]  \\    
\hline                        
\#1& $\langle l=0,1 \rangle$  & 0.26$\pm$0.35 &  1.35$\pm$0.36\\     
\#1& $\langle l=0 \rangle$ & 0.87$\pm$0.64 & 2.26$\pm$0.61\\
\#1& $\langle l=1 \rangle$ & 0.01$\pm$0.42 & 0.87$\pm$0.44\\
\hline    
\#2& $\langle l=0,1,2 \rangle$ & 0.51$\pm$0.41 &  1.23$\pm$0.43\\     
\#2& $\langle \mbox{even } l\rangle$ & 1.15$\pm$0.67 & 2.34$\pm$0.82\\ 
\#2& $\langle \mbox{odd } l\rangle$  & 0.12$\pm$0.53 & 0.80$\pm$0.51\\ 
\hline
\#3& $\langle {\rm all}$ $l \rangle$ & 0.19$\pm$0.12& 1.32$\pm$0.18 \\   
\#3& $\langle {\rm even}$ $l \rangle$ & 0.64$\pm$0.17& 1.70$\pm$0.55 \\   
\#3& $\langle {\rm odd}$ $l \rangle$ & $-0.05\pm0.15$& 0.83$\pm$0.17 \\    
\hline             
\end{tabular}
\tablefoot{The methods $\#$1 and $\#$2 correspond, respectively, to the local and global mode peak-fitting analyses (Sect.~\ref{sec:fitting}), and the method $\#$3 corresponds to the cross-correlation analysis (Sect.~\ref{sec:cross}).}
\end{table}

\section{Summary}
The p-mode frequencies and amplitudes of the solar-like oscillating star HD~49933 were observed to vary in time with the magnetic activity of the star \citep{garcia10}. We studied here the frequency dependence of the p-mode frequency shift by analyzing the same observations as \citet{benomar09} and \citet{garcia10} collected by the CoRoT satellite. The frequency shifts were obtained independently through the analysis of the mode frequencies extracted by both a local and a global peak-fitting procedures,  and through a cross-correlation method. We showed that the frequency shifts in HD~49933 increase with frequency reaching a maximum of about 2 $\mu$Hz at 2100 $\mu$Hz.
We also inferred that the observed variations in mode frequencies are related to changes in the outer layers of HD~49933 due to its magnetic activity, as the activity-frequency relationship with frequency is analogous in shape to the one observed along the 11-year solar cycle. HD~49933 is thus the second star after the Sun for which the frequency dependence of the p-mode frequency shifts with magnetic activity has been measured.

\begin{acknowledgements}
The authors thank the referee of this paper for providing useful comments that improved the presentation. The CoRoT space mission has been developed and is operated by CNES, with contributions from Austria, Belgium, Brazil, ESA (RSSD and Science Program), Germany and Spain. DS acknowledges the support from the Spanish National Research Plan (grant PNAyA2007-62650) and from CNES. This work was supported by the CNES/GOLF grant and the ``Programme National de Physique Stellaire'' at SAp/CEA-Saclay.
The National Center for Atmospheric Research is sponsored by the U.S. National Science Foundation.
\end{acknowledgements}


\end{document}